\newcommand{\PSD}{\hat{\psi}^\dagger}
\newcommand{\PSI}{\hat{\psi}}
\newcommand{\rv}{({\bf r})}

\documentstyle[twocolumn,epsf]{jpsj}

\def\runtitle
{
Bose-Einstein Condensation in a Confined Geometry
}
\def\runauthor
{
Tomoya  Isoshima and 
Kazushige Machida
}

\title
{
Bose-Einstein Condensation in a Confined Geometry with  and without a Vortex
}
\author
{
Tomoya {\sc Isoshima}\footnote{E-mail: tomoya@mp.okayama-u.ac.jp} and 
Kazushige {\sc Machida}\footnote{E-mail: machida@mp.okayama-u.ac.jp} }
\inst
{
Department of Physics, Okayama University, Okayama 700
}
\recdate
{
July 11, 1997
}
\abst
{Various widely-used mean-field type theories
for a dilute Bose gas are critically examined
in the light of the recent discovery of Bose-Einstein
condensation of atomic gases in a confined geometry.
By numerically solving the mean-field equations within the framework of
the Bogoliubov approximation both 
stationary non-uniform case and the vortex case under rotation 
in a cylindrically symmetric vessel are investigated.
We obtain spatial structures of condensate, non-condensate, anomalous
correlation.
The low lying excitation spectra, the local density of states
and the circulating current density in a vortex
corresponding to various levels of mean-field theories
are predicted. 
}

\kword
{
Bose-Einstein condensation, alkali atom gas, vortex,
Hartree-Fock Bogoliubov approximation, Gross Pitaevskii equation,
Bogoliubov equation
}

\begin{document}

\maketitle

\section{Introduction}
Much attention has been focused on many-body dilute Bose systems
since the discovery of Bose-Einstein condensation (BEC) 
in alkali atom gases at ultra-low temperatures around
$O(\mu K)\sim O(nK)~$\cite{cornell,hulet,ketterle}.
The current experiments~\cite{myatt} performed on atomic gases
such as sodium and rubidium in confined geometries trapped by
a magnetic field and/or optical method.
Furious experimental and theoretical investigations are
now ongoing and rapidly developing.
Since these atom gases are dilute Bose systems,
allowing us to model these as weakly interacting
many-body Bose systems,
it is regarded as realization of ideal Bose-Einstein condensation.
Although superfluid $^4$He has been considered as the first BEC,
the mutual repulsive interaction 
is rather strong where the condensate fraction  only $\sim$10$\%$ of the total.
There is little chance to directly apply microscopic theory to it.

The microscopic theoretical work on BEC started with
Bogoliubov~\cite{bogoliubov},
followed by several important progresses, 
such as Gross~\cite{gross}, Pitaevskii~\cite{pitaevskii},
Iordanskii~\cite{iordanskii}, and Fetter~\cite{fetter}.
They treat an infinite system.
These mean-field theories~\cite{review} are particularly suited for
treating an spatially non-uniform system which is the case of
the present gaseous BEC systems trapped optically or magnetically
in a restricted geometry. 
Therefore, the current theoretical studies~\cite{meanfield} mainly
focus on examining these mean field theories at various stages of
the employed approximation for dilute Bose gases trapped by
the usually harmonic potential in order to extract the spatial structures
of the condensate and non-condensate and low-lying collective modes.
Agreement between these mean field theories and experiments is generally
fairly good so far, encouraging us to go along this line.

The purposes of this paper are two-fold:
One is to compare various approximate treatments:
(1) the Gross-Pitaevskii approximation (GP), which considers only
the condensate,
(2) the Bogoliubov approximation (BA), which takes into account 
the non-condensate
too, but not self-consistent,
(3) the Popov approximation (PA), which neglects the anomalous correlation and
(4) the Hartree-Fock Bogoliubov approximation (HFB).
(3) and (4) take into account the condensate and non-condensate
self-consistently.
We obtain a complete self-consistent solution numerically for these four cases
on an equal footing. This has not been done before. This is particularly true
for HFB whose detailed study has not been
performed. 
These various approximate theories reveal both merits and demerits.
None is complete and satisfactory. 
Nevertheless we may gain some insights to general properties of
trapped BEC systems.

The second purpose, which is the main purpose,
is to know various properties of the quantized vortex structure
in rotating BEC systems since microscopic treatment of the vortex 
has been lacking so far except for a few attempts based
on GP~\cite{gpvortex} for present BEC systems.
We are particularly interested in understanding
(A) the spatial profiles of the condensate, non-condensate and
anomalous correlation,
(B) the dispersion relation of the  excitation spectrum of the system,
and (C) the low-lying excitations localized around a vortex core in detail,
which is known as the Kelvin mode~\cite{donnelly}.
These features should be compared with those in vortices in a superconductor
where extensive studies have been compiled~\cite{hohen}.
This is particularly true for the localized excitations near a 
vortex core studied by Caroli {\it et al.}~\cite{caroli}.
The HFB theory just corresponds to the so-called 
Bogoliubov de Gennes theory (BdG), which is widely used
to treat various spatially non-uniform superconductors 
having interface or vortex.
In particular, Gygi and Schl\"uter~\cite{gygi} have done a detailed
investigation on an isolated vortex, providing a 
fairly complete and clear picture of microscopic structures
for quantized vortex in a s-wave superconductor. 

We have been motivated by their impressive work
and considered that a similar work should be done in a BEC system
because vortex in BEC might be eventually observed
and the vortex core structure and associated low-lying
excitations may be probed experimentally.
The characteristic core radius is estimated as the 
coherence length $\sim 0.1\mu$m which 
contrasted with a few $\AA$ in superfluid $^4$He.
Thus there is a good chance to investigate
the detailed core structure of the present BEC systems.
In a superconductor scanning electron microscopy by 
Hess {\it et al.}~\cite{hess} directly gives vivid spatially
resolved images of the low-lying excitations near a vortex core.

It turns out that although the full self-consistent HFB
solution obtained here shows some fundamental
unsatisfactory caveats of the gapfull excitation 
which are not in BdG, we believe that the present study 
is of still some use, because the overall features of
the resulting vortex picture are expected to be 
independent of the approximations.
These physical quantities obtained here may be directly observable
once the experiment succeeds in creating a vortex. 

In his series of papers Fetter~\cite{fetter} investigates the vortex structure
in an imperfect Bose gas within the Bogoliubov approximation,
giving an approximate analytic treatment for an 
infinite system:
Although he succeeded in deriving asymptotic behaviors of several
physical quantities, such as the non-condensate fraction at the core using 
the approximately obtained eigenfunctions on the basis of the work
by Pitaevskii~\cite{pitaevskii} and Iordanskii~\cite{iordanskii}
who solve the Bogoliubov equation for the particular 
angular momentum states: $q_{\theta}=-1$ and $q_{\theta}=1$
respectively ($q_{\theta}$ is defined shortly).
Hence the problem with the rotating BEC in the Bogoliubov
theory is not solved completely even for an infinite system.
Thus no one has analyzed it for a finite trapped system.

In next Section a brief description of the 
Hartree-Fock Bogoliubov theory (HFB) is given.
In this process various approximate treatments mentioned above are introduced.
We explain how to solve a set of the HFB equations numerically
in a self-consistent manner, following the method devised by
Gygi and Schl\"uter~\cite{gygi}. Here we consider a cylindrically
symmetric case with a rigid wall and choose parameters appropriate to
present BEC systems for $^{23}$Na and $^{87}$Rb.
The results for stationary non-uniform case are presented in \S 3.
The vortex structure is discussed in details within HFB which turns
out to be only stable theory among the above theories except for GP when
the system rotates around the symmetry axis in \S 4.
The final section is devoted to discussions and summary.

\section{Formulation and Numerical Procedure}
\subsection{Various Approximations}

We start with the following Hamiltonian in which 
Bose particles interact with a two-body potential $V_{\rm int}$:
\begin{eqnarray}
   \hat{\rm H} &=& \int {\rm d}{\bf r}
   \hat{\Psi}^{\dagger}\rv  
   \left\{
     - \frac{\hbar ^2 \nabla ^2}{2m} + V_{\rm ext}\rv - \mu
   \right\}
   \hat{\Psi}\rv \nonumber
\\& &
   + \frac{1}{2} 
   \!\! \int \!\! {\rm d}{\bf r}_1 \!\!\! \int \!\! {\rm d}{\bf r}_2
   \hat{\Psi}^{\dagger} ({\bf r}_1) \hat{\Psi}^{\dagger} ({\bf r}_2) 
   V_{{\rm int}}( {\bf r}_1 - {\bf r}_2 )
   \hat{\Psi}({\bf r}_2)    \hat{\Psi}({\bf r}_1)  \label{eq:h1}
\end{eqnarray}
where the chemical potential $\mu$ is introduced
to fix the particle number
and $V_{{\rm ext}}\rv$ is the confining potential.
In order to describe the Bose condensation, we assume that the field operator
$\hat{\Psi}$ is decomposed into
\begin{equation}
   \hat{\Psi} =
   \PSI\rv + \phi\rv \label{eq:deco}
\end{equation}
where the ground state average is given by
\begin{equation}
   \langle \hat{\Psi}\rv \rangle = \phi\rv.
\end{equation}
A c-number $\phi\rv$ corresponds to the condensate wave function
and $\PSI\rv$ is a q-number describing the non-condensate.
The two-body interaction $V_{\rm int}({\bf r}_1 - {\bf r}_2)$ is assumed to
be $g\delta({\bf r}_1 - {\bf r}_2)$ with $g$ being a positive (repulsive)
constant
proportional to the s-wave scattering length $a$, namely
$g=4\pi \hbar^2a/ m$ ($m$ the particle mass).
Substituting the above decomposition (\ref{eq:deco}) in (\ref{eq:h1}),
we obtain
\begin{eqnarray}
  \hat{{\rm H}}
  &=&
  \int {\rm d}{\bf r}
  \Bigl[
    \phi^*\rv h\rv\phi\rv +{1\over 2}g|\phi|^4    \nonumber
\\& &
    +\PSD\rv \left\{
      h\rv + g|\phi\rv|^{2}
    \right\}\phi\rv 
    +{\rm h.c.}       \nonumber
\\ & &
    +\PSD\rv \left\{
      h\rv
    +2g|\phi\rv |^{2}  
    \right\} \PSI\rv    \nonumber
\\ & &
    +\frac{g}{2}\PSD\rv \PSD\rv \phi\rv \phi\rv  +{\rm h.c.}   \nonumber
\\ &&
    +g\PSD\rv \PSI\rv \PSI\rv \phi^*\rv   +{\rm h.c.}     \nonumber
\\ &&
    +\frac{g}{2}\PSD\rv \PSD\rv \PSI\rv \PSI\rv 
  \Bigr]
\end{eqnarray}
with
\begin{equation}
   h\rv \equiv
   -\frac{\hbar^2\nabla^2}{2m} + V_{\rm ext}\rv - \mu \label{eq:h}
\end{equation}
one-body Hamiltonian.
Let us introduce the variational parameters: 
the non-condensate density
$\rho\rv = \langle \PSD\rv \PSI\rv \rangle$ 
and the anomalous correlation
$\Delta\rv = \langle \PSI\rv \PSI\rv \rangle$
and approximate as 
\begin{eqnarray}
  \PSD \PSI \PSI
  &=&
  2\PSI \rho + \PSD \Delta
\\
  \PSD \PSD \PSI \PSI
  &=&
  \PSD \PSD \Delta + \PSI \PSI \Delta^* + 4\PSD \PSI \rho. 
\end{eqnarray}
Then, $\hat{\rm H}$ is rewritten as
\begin{eqnarray}
  \hat{\rm H}
  &=&
  \int \!{\rm d}{\bf r}\Bigl[
    \Bigl\{
      \phi^*\rv h\rv \phi\rv +{1\over 2}g|\phi|^4
    \Bigr\}                      \nonumber
\\&&
    +\PSD\rv \bigl\{
      h\rv + g|\phi\rv|^{2} +2g\rho\rv 
    \bigr\} \phi\rv
    +{\rm h.c.}                        \nonumber
\\&&
    +\PSD\rv g\Delta\rv \phi^*\rv 
    +{\rm h.c.}                        \nonumber
\\&&
    +\PSD\rv \left\{
      h\rv + 2g|\phi\rv|^{2} + 2g\rho\rv
    \right\} \PSI\rv               \nonumber 
\\&&
    +\frac{g}{2} \PSD\rv \left\{
      \Delta\rv + \phi^{2}\rv
    \right\} \PSD\rv  + {\rm h.c.}
  \Bigr].
\end{eqnarray}
In order to diagonalize this Hamiltonian, the following Bogoliubov
transformation is employed, namely, 
$\PSI\rv $ is written in terms of 
the creation and annihilation operators $\eta _q$ and $\eta _q^{\dagger}$ 
and the non-condensate wave functions $u_q\rv $ and $v_q\rv $ as
\begin{equation}
   \PSI\rv  = \sum _{{\it q}}
      \left[
         u_q\rv \eta _q - v_q^*\rv \eta _q^{\dagger}
      \right]   \label{eq:psieta}
\end{equation}
where $q$ denotes the quantum number.
This leads to the diagonalized form:
\begin{equation}
   \hat{\rm H} = {\rm E}_0 + \sum_qE_q\eta _q^{\dagger}\eta _q.  
\end{equation}
The condition that the first order term in
$\PSI\rv $ vanish yields
\begin{equation}
  \{ 
    h\rv + g|\phi\rv|^{2} + 2g\rho\rv
  \} \phi\rv 
   + g\Delta\rv \phi^*\rv  = 0.  \label{eq:gp}
\end{equation}
When $\rho\rv $ and $\Delta\rv $ are made zero, %%%%%%<made to zero,> 
it reduces to the Gross-Pitaevskii (GP) 
equation:
\begin{equation}
   h\rv \phi\rv 
   + g\phi^*\rv \phi\rv \phi\rv 
    = 0,  \label{eq:gpo}
\end{equation}
which is a non-linear Schr\"odinger type equation.

The condition that the Hamiltonian be diagonalized gives rise to 
the following set of eigenvalue equations for $u_q\rv $ and
$v_q\rv $ with the eigenvalue $E_q$:
\begin{eqnarray}
   \left[
      h\rv + 2g\{ \rho\rv + |\phi\rv|^2 \}
   \right] u_q\rv   \nonumber
&&\\
   -g\left[
      \Delta\rv  + \phi ^2\rv 
   \right] v_q\rv 
   &=& E _qu_q\rv    \label{eq:bg1}
\\
   \left[
      h\rv  + 2g\{\rho\rv  + |\phi\rv|^2 \}
   \right] v_q\rv    \nonumber
&&\\
   -g\left[
      \Delta^*\rv  + \phi ^{*2}\rv 
   \right] u_q\rv 
   &=& -E_qv_q\rv .   \label{eq:bg2}
\end{eqnarray}
The eigenvectors
$u_q\rv $ and  $v_q\rv $ must satisfy the  normalization condition:
\begin{equation}
   \int \left\{
      u_p^*\rv u_q\rv  - v_p^*\rv v_q\rv 
   \right\} {\rm d}{\bf r}
   = \delta _{p,q}. \label{eq:nor}
\end{equation}
The variational parameters
$\rho\rv $ and  $\Delta\rv $ are determined self-consistently by 
\begin{eqnarray*}
  \rho\rv
  &=&
  \langle \PSD \PSI \rangle
\\
  &=&
  \sum_{\it q}
    (|u_q\rv|^{2} + |v_q\rv|^{2})f(E_q)
    +\sum_{\it q} |v_q\rv|^{2}
\\
  \Delta\rv 
  &=&
  \langle \PSI \PSI \rangle
\\
  &=&
  -\sum_{{\it q}}
     2u_q\rv v_q^*\rv f(E_q)
     -\sum_{\it q} u_q\rv v_q^*\rv  
\end{eqnarray*}
where $f(E)$ is the Bose distribution function.
At absolute zero temperature, which we consider from now on, these become
\begin{eqnarray}
   \rho\rv  &=& \sum_{{\it q}} v_q^*\rv v_q\rv  \label{eq:rho}
\\
   \Delta\rv  &=& -\sum_{{\it q}} u_q\rv v_q^*\rv . \label{eq:del}
\end{eqnarray}
Equations (\ref{eq:gp}), (\ref{eq:bg1}), (\ref{eq:bg2}),
(\ref{eq:rho}) and (\ref{eq:del}) constitute a complete set of the 
self-consistent equations for the Hatree-Fock Bogoliubov theory (HFB).
If $\rho\rv $ and $\Delta\rv $ are made to zero, we recover the Bogoliubov
theory (BA).
The Popov approximation (PA) is the case of $\Delta\rv =0$ in HFB.
In the followings, we perform these four cases, namely,
GP, BA, PA and HFB in the equal footing to 
comparatively study similarities and differences.

The expectation value of the particle number density is
given as
\begin{eqnarray}
   \langle \hat{\rm n}\rv \rangle = |\phi\rv |^2 + \rho\rv ,
\end{eqnarray}
that is, the total density consists of the condensate part $\phi\rv $
and the non-condensate part $\rho\rv $.
The particle current density is calculated,
\begin{eqnarray}
  {\bf j}\rv  
  &=&
  \frac{\hbar}{2m{\rm i}}\{
    \phi^*\rv \nabla\phi\rv
    -\phi\rv \nabla\phi^*\rv
  \}     \nonumber
\\&&
  +\frac{\hbar}{2m{\rm i}}\langle 
    \PSD\rv \cdot \nabla \PSI\rv  
    -\nabla \PSD\rv \cdot \PSI\rv 
  \rangle.
\end{eqnarray}
%%%%%%%%%%%%%%%%%%%%
\subsection{Numerical Procedure}

For later convenience, we introduce the following non-dimensional quantities:
In terms of the mass $m$ and the average particle number density $n_a$,
the various densities and the length are  scaled by  $n_a$ and
$\xi_a \equiv {\hbar ^2}/{\sqrt{2mn_ag}}$ respectively.
$\xi_a $ is the coherence length of the condensate given
by solving the above GP equation (\ref{eq:gpo}). The energy 
is scaled by $n_ag$.
We define the quantities:
${\bf r}^\prime \equiv  \frac{1}{\xi _a}{\bf r}$,
$\phi^\prime({\bf r}^\prime)  \equiv \frac{1}{\sqrt{n_a}}\phi\rv$, 
$\rho^\prime({\bf r}^\prime) \equiv  \frac{1}{n_a}\rho\rv$,
$\Delta^\prime({\bf r}^\prime)\equiv \frac{1}{n_a}\Delta\rv$,
$u^\prime({\bf r}^\prime) \equiv  \frac{1}{\sqrt{n_a}}u\rv$,
$v^\prime({\bf r}^\prime) \equiv  \frac{1}{\sqrt{n_a}}v\rv$,
$E ^\prime _q\equiv \frac{1}{n_ag}E_q$.
The system is now characterized by 
$n_a\xi_a^3$ and the system volume normalized by $\xi_a$.
From now on, we suppress primes $\prime $ in these newly defined quantities.
Note that by increasing the average density 
$n_a$ the effective interaction gets stronger, namely
the tuning of the interaction by controlling an external parameter $n_a$
is an interesting and important aspect of the BEC system.

We now consider a cylindrically symmetric system
which is characterized by the radius $R$ and the height $L$.
 We impose the boundary conditions that in terms of the 
cylindrical coordinate:  ${\bf r} = (r,\theta ,z)$
the all wave functions vanish at the wall $r=R$ and
the periodic boundary condition along the $z$-axis.
When a vortex line passes through the center of the cylinder
the condensate wave function $\phi\rv $ is expresses as
\begin{equation}
  \phi(r,\theta ,z) = \phi(r) e^{{\rm i}w\theta } 
\end{equation}
where $\phi(r) $ is a real function and 
$w$ is the winding number. $w=0$ 
corresponds to non-vortex case 
and $w=1$ to the vortex case.
The $w\ge 2$ case is not considered here because
this state is energetically unstable.
The non-condensate density $\rho $ is a real function, depending  only on $r$
that is, $ \rho (r,\theta ,z)  = \rho(r)$.
It is seen from eq.\ (\ref{eq:gp}) that
the anomalous correlation $\Delta $ has the phase with 2$\theta $, thus
 $\Delta (r,\theta ,z) = \Delta(r) e^{2{\rm i}w\theta }$.
It is also seen from eqs.\ (13) and (14) that the phases of 
$u_q\rv $ and $v_q\rv $ are written as
\begin{eqnarray}
u_{\bf q}\rv &=& u_{\bf q}(r)e^{{\rm i}q_zz}e^{{\rm i}(q_{\theta } + w)\theta}\\
v_{\bf q}\rv &=& v_{\bf q}(r)e^{{\rm i}q_zz}e^{{\rm i}(q_{\theta } - w)\theta }.
\end{eqnarray}
The set of the quantum numbers ${\bf q}$ in (\ref{eq:psieta})
is described by $(q_r, q_{\theta } , q_z)$ where
$q_r = 1, 2, 3, \cdots $, $q_\theta = 0, \pm 1, \pm 2, \cdots$, 
$q_z = 0, \pm 2\pi /L, \pm 4\pi /L, \cdots$.

The functions  $u_q(r)$  and $v_q(r)$ are expanded in terms of 
\begin{equation}
   \varphi _{\mu }^{(i)}(r)
   \equiv \frac{\sqrt{2}}{J_{|\mu| +1}\left(\alpha _{\mu}^{(i)}\right)}
   J_{|\mu|}\left(\alpha _{\mu }^{(i)}\frac{r}{R}\right)
\end{equation}
as
\begin{eqnarray}
   u_q(r)
   &=&
   \sum_i c_q^{(i)}\varphi_{q_{\theta }+w}^{(i)}(r) \label{eq:phase:u}
\\
   v_q(r)
   &=&
   \sum_i d_q^{(i)}\varphi_{q_{\theta }-w}^{(i)}(r) \label{eq:phase:v}
\end{eqnarray}
where $J_{\mu}(r)$ is the Bessel function of $\mu$-th order and
$\alpha_{\mu}^{(i)}$ denotes $i$-th zero of $J_{\mu}$.

Then, eq.\ (\ref{eq:gp})  becomes
\begin{eqnarray}
  -\left\{
    \frac{{\rm d}^2\phi }{{\rm d}r^2}
    +\frac{1}{r} \frac{{\rm d}\phi }{{\rm d}r} - \frac{w^2}{r^2}\phi
  \right\}  \nonumber
&&\\
  +\left\{
    V_{\rm ext} - \mu + \phi^2 + 2\rho + \Delta
  \right\}\phi
  &=& 0.  \label{eq:c:gp}
\end{eqnarray}
The eigenvalue problem of eqs.\ (\ref{eq:bg1}) and (\ref{eq:bg2})
reduces to diagonalizing 
the matrix:
\begin{equation}
  \left(\!\!\!\!
    \begin{array}{cc}
      A_{i,j}(q_{\theta }+w,q_z) & -B_{i,j}(q_{\theta },w)   \\
      & \\
      B_{i,j}^{\rm T}(q_{\theta },w) & -A_{i,j}(q_{\theta }-w,q_z)
    \end{array}
  \!\!\!\!\right)\!\!\!
  \left(\!\!\!\!
    \begin{array}{c}
      c_{\bf q}^{(1)}   \\
      c_{\bf q}^{(2)}   \\
      \vdots    \\
      d_{\bf q}^{(1)}   \\
      d_{\bf q}^{(2)}   \\
      \vdots
    \end{array}
  \!\!\!\!\right)
  \! = \!
  E_{\bf q}  \!\!
  \left(\!\!\!\!
    \begin{array}{c}
      c_{\bf q}^{(1)}   \\
      c_{\bf q}^{(2)}   \\
      \vdots    \\
      d_{\bf q}^{(1)}   \\
      d_{\bf q}^{(2)}   \\
      \vdots
    \end{array}
  \!\!\!\!\right) \label{eq:c:mat}
\end{equation}
\begin{eqnarray}
  A_{i,j}(\mu ,q_z) 
  &=&
  \left\{
    \left(\frac{\alpha_{\mu}^{(j)}}{R} \right)^2
    + q_z^2 - \mu 
  \right\}\delta _{i,j}  \nonumber
\\&&
  +\int_0^R \!\!
    \left( V_{\rm ext} + 2(\phi^2 + \rho) \right)
    \varphi _{\mu }^{(i)}\varphi _{\mu }^{(j)}
  r\,{\rm d}r 
\\ \label{eq:c:mat:a}
  B_{i,j}(q_{\theta },w) 
  &=&
  \int _0^R
    \left( \Delta + \phi ^2 \right)
    \varphi _{q_{\theta }+w}^{(i)}\varphi _{q_{\theta }-w}^{(j)} 
  r\,{\rm d}r \label{eq:c:mat:b}
\end{eqnarray}
where ${\bf q}=(q_r, q_\theta, q_z)$.
The normalization condition (\ref{eq:nor}) of the eigenvectors 
is rewritten in terms of $c^{(i)}_{\bf q}$  and $d^{(i)}_{\bf q}$ as 
\begin{equation}
   n_a\xi_a^3 \sum _i \left\{
      c_{\bf p}^{(i)}c_{\bf q}^{(i)} - d_{\bf p}^{(i)}d_{\bf q}^{(i)}
   \right\}
   = \frac{1}{2\pi L}\delta _{{\bf p},{\bf q}}.
\end{equation}  \label{eq:c:nor}
From eqs.\ (\ref{eq:rho}) and (\ref{eq:del}) $\rho(r) $ and  
$\Delta(r) $ are determined by
\begin{eqnarray}
   \rho(r)    &=& \sum_{{\bf q},i} 
      (d_{\bf q}^{(i)}
      \varphi _{q_{\theta }-w}^{(i)})^2 
      \label{eq:c:rho}
\\
   \Delta(r) &=& - \sum_{{\bf q},i,j}
      c_{\bf q}^{(i)}d_{\bf q}^{(j)} 
      \varphi _{q_{\theta }+w}^{(i)}\varphi _{q_{\theta }-w}^{(j)}.
            \label{eq:c:del}
\end{eqnarray}

The iterative calculations of 
eqs.\ (\ref{eq:c:gp}), (\ref{eq:c:mat}),
(\ref{eq:c:rho}), and (\ref{eq:c:del}) yield a convergent
self-consistent solution.
It is not self-evident a priori that eq.\ (\ref{eq:c:mat}) gives real eigenvalues
because the Hamiltonian matrix in eq.\ (\ref{eq:c:mat}) is not symmetric.
Note that since $q_\theta$ is the angular momentum and
$q_z = 2\pi n/L$ ($n$ integer) is the wave number along the $z$-axis, both being 
good quantum numbers, the eigenvalue equation (\ref{eq:c:mat}) 
is decomposed into each $q_\theta$ and $q_z$.
Each block-diagonal eigenvalue equation
gives rise to the quantum number $ q_r$ along the radial direction $r$.

\subsection{Calculated System}

As mentioned, our system is characterized by the two parameters:
the reduced density $n_a\xi_a^3$ and the system size $R$ and $L$ 
(in unites of $\xi_a$).
The interaction strength 
characterized by the scattering length $a$ is absorbed in the reduced density.
We set $n_a\xi_a^3=1.0$, $R=15$, and $L=30$
throughout this paper. We are treating 2.12$\times10^4$ atoms.
Since the scattering length of Rb is known as $a$=5.8nm, 
the energy scale $n_ag\sim10^{-30}$J.
These numbers are compared with the experiment
by Anderson {\it et al.}~\cite{cornell}.
The radius of the condensate
$\sim10\mu$m, the trapping potential energy $\sim10^{-28}$J
and the total number of atoms $\sim$10$^3$.
Similar set of the numbers are obtained for the Na experiment~\cite{ketterle}.

We have employed the two methods for numerical calculation:
the momentum cutoff method
and the energy cutoff method to check our numerical accuracy.
In the former, the computation is limited to 
certain finite quantum numbers for $q_r$, 
$q_\theta$, and $q_z$.
We have examined several cases where
-50$\le$$q_\theta$, $q_z$$\le$+50 and -75$\le$$q_\theta$, $q_z$$\le$+75,
keeping the maximum number of $q_r\sim 50$ fixed.
In view of the tendency of the obtained solutions,
the last case is satisfactory, thus the total number of the 
treated eigenfunctions $\sim 500000$
because the size dependence on $q_{\theta}$ and $q_z$ almost ceases stopping.
In the energy cutoff method where the calculation terminates when the obtained
eigenvalues exceeds a certain value, typically, 100, and 200 in the units
of $n_ag$.
The last case which treats $\sim 500000$ eigenfunctions seems best
in view of the expected behavior of the solutions.
Note that the anomalous correlation $\Delta$ is most dependent
on this cutoff condition while $\rho$ is 
relatively insensitive.
 Comparing the two methods,
the energy cutoff method is better than the momentum cutoff method,
thus we show the results employed this method below.

\section{Stationary non-uniform  case}

In order to examine the accuracy of the numerical computations and
to know the properties of BEC in the stationary state,
we first consider non-vortex Bose system confined in a cylindrical vessel.
The corresponding 
infinite system is analyzed by, for example, Fetter~\cite{fetter}
within the Bogoliubov approximation (BA).

The condensate $\phi\rv $ is shown in Fig.\ 1
where GP and BA give the same result for this quantity,
since BA neglects the non-condensate and the anomalous correlation.
It is seen that PA and HFB give nearly same results as that in GP (or BA)
because the absolute values of $\rho$ and $\Delta$
are very small for our parameter selected. Thus this is
not the general conclusion. The condensate fraction could
decrease as $g$ or $n_a$ increases.
The overall behaviors of these results are quite understandable;
The condensate changes only near the wall, whose characteristic
length is the coherent length $\xi_a$
introduced before.
The expected flatness in the bulk region around the center indicates
the reliability of our numerical calculations.

%-------------------------------------------
\begin{figure}
\epsfxsize=7.5cm
\epsfbox{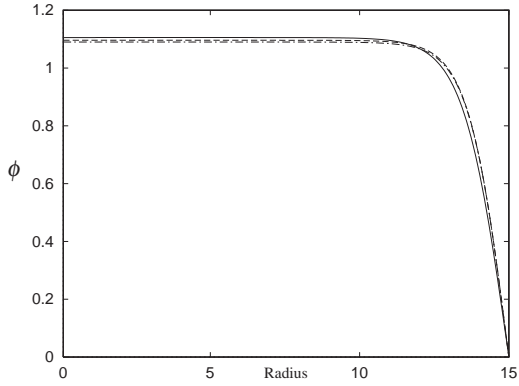}
\caption{
%Fig.1: 
Spatial variation of the condensate $\phi(r)$ as a function of $r$ for
HFB (bold line), PA (dot dashed line), and GP and BA (dashed line).
}
\label{fig:1}
\end{figure}
%-------------------------------------------

The spatial profiles of the non-condensate $\rho\rv $
for various approximations are displayed in Fig.\ 2.
$\rho\rv $ in BA is calculated by using eq.\ (16)
where $u(r)$ and $v(r)$ are a solution in BA
and is shown for reference.
While BA and PA give a similar variation, the magnitude of $\rho\rv $ in HFB
is almost halved. This difference 
may be related to $\Delta$ in HFB. It is noted that 
$\rho(0)\sim 0.014$ in BA and PA compared with the analytic expression at $T$=0 
\begin{eqnarray}
\rho={1\over 6\pi^2{\sqrt2}\xi^3 n_0}\sim0.012{1\over\xi^3n_0}
\end{eqnarray}
by Fetter~\cite{fetter} for an infinite system
($n_0$ the average number density and $\xi=\hbar/\sqrt{2mn_0g}$) 
whose number $\rho=0.012$ in the present case ($\xi^3 n_0$=1.0).
The small difference comes  from the system size (finite vs infinite).
In fact  in our calculations the limiting value $\rho=0.012$ 
are slowly recovered as the system size increases.
The major spatial variation only occurs at the
wall whose length is an order of $\xi_a$.
Note from Fig.\ 2 that $\rho$ vanishes quadratically instead of linearly in 
$\phi(r)$.

%-------------------------------------------
\begin{figure}
\epsfxsize=7.5cm
\epsfbox{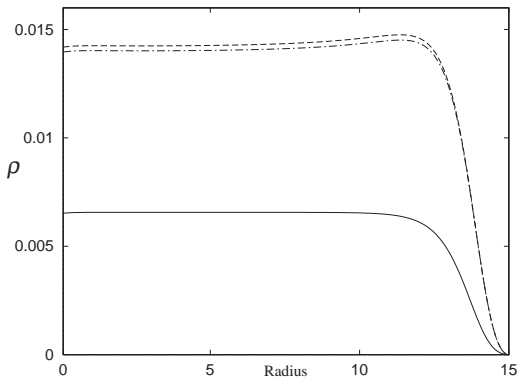}
\caption{
Spatial variation of the non-condensate $\rho(r)$ as a function of
$r$ for HFB (bold line), PA (dot dashed line),
and BA (dashed line).
}
\label{fig:2}
\end{figure}
%-------------------------------------------

The anomalous correlation $\Delta(r)$ is
shown in Fig.\ 3 where BA and HFB are compared.
The amplitude $\Delta(r)$ in HFB at the center are rather
large and negative, which affects mainly on $\rho(r)$,
namely, the large difference of $\rho(r)$ between 
BA and PA, and HFB comes from the absence or presence of 
$\Delta(r)$. It is noted also that $|\Delta(r)|$ in the bulk region
sensitively depends on the calculated system size which
is contrasted with other quantities such as $\phi(r)$ or $\rho(r)$.
It should be noticed also that as mentioned above 
the expected flat behavior far from the wall is 
reproduced only when the enough number of the eigenfunctions
are taken account in the numerical computation, otherwise 
this particular quantity $\Delta(r)$ often fails to 
exhibit the expected flat
behavior.

%-------------------------------------------
\begin{figure}
\epsfxsize=7.5cm
\epsfbox{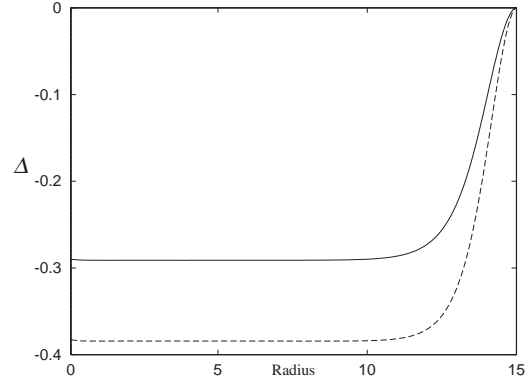}
\caption{
Spatial variation of the anomalous correlation $\Delta(r)$ as a
function of $r$ for HFB (bold line) and BA (dashed line).
}
%%%\label{fig:3}
\end{figure}
%-------------------------------------------

In Figs.\ 4(a), (b), and (c) where the eigenvalues $E({\bf q})$ are plotted as
functions of the quantum numbers of $q_{\theta}$ and $q_z$,
the excitation spectra are shown for BA and HFB. 
The spectrum in PA which is not shown here is almost identical to that in BA.
Indeed, the excitation spectrum in PA and BA
is gapless while HFB 
is gapful as expected~\cite{hohenberg}.
In BA the so-called Bogoliubov spectrum~\cite{fetter} 
$E^B({\bf k})$ is known to be given by 
\begin{equation}
  E^B({\bf k})=\sqrt{\epsilon^2_k+2\epsilon _kn_0g}
\end{equation}
where $\epsilon _k={\hbar^2{\bf k}^2\over 2m}$.
In the long wavelength limit it reduces to
\begin{eqnarray}
  E^B({\bf k})\sim\hbar ck \ \ \ \ \ {\rm for}\ \ \ \ k\rightarrow 0
\end{eqnarray}
where $c=\sqrt{{n_0g\over m}}$.
The low-lying excitations are known to be exhausted by phonons
($c$ is the sound velocity)~\cite{pines}.
In the short wavelength limit it becomes
\begin{eqnarray}
  E^B({\bf k})\sim{\hbar^2{\bf k}^2\over 2m} \ \ 
  {\rm for}\ \ k\rightarrow\infty.
\end{eqnarray}
where the excitations are individual one-particle excitation.
These expected behaviors in BA are well reproduced by the present calculations;
As seen from Figs.\ 4(a) and 4(b),
the dispersion relations in BA and PA are linear 
in $q\rightarrow0$ and quadratic in $q\rightarrow\infty$
where $k_z={2\pi\over L}q_z$.
In fact, the theoretical lines of eqs.\ (35) and (36) drawn in Figs.\ 4(a) and
4(b) show good fits to the numerical results
without any fitting parameter, proving the reliability
of our calculations.
On the other hand, as seen from Fig.\ 4(c) 
HFB is a gapful theory. This dispersion relation is quadratic in
both $q_{\theta}$ and $q_z$. The magnitude of this gap 
is an order of $n_a g$.

%-------------------------------------------
\begin{figure}
\epsfxsize=8cm
\epsfbox{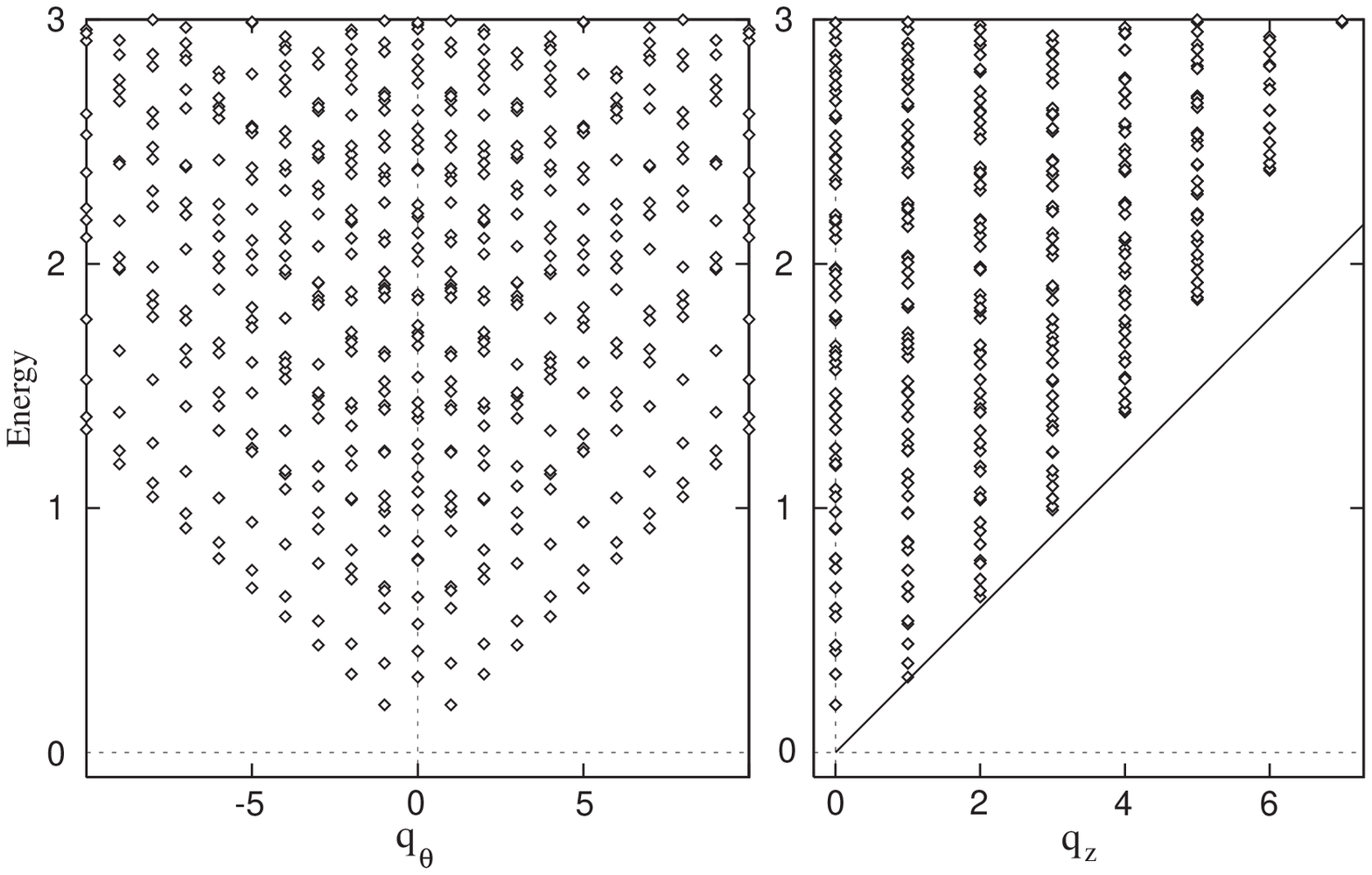}

\epsfxsize=6cm
\epsfbox{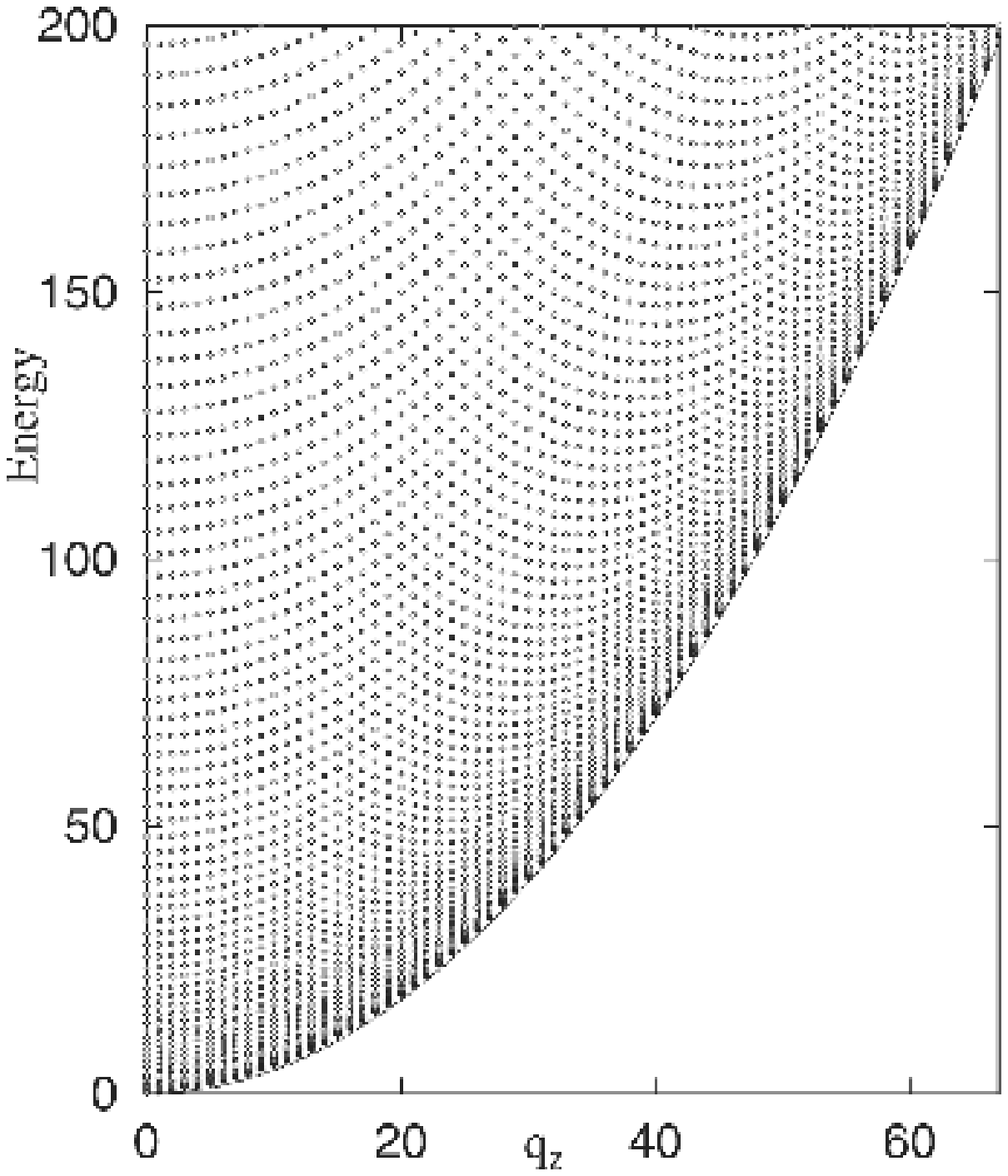}

\epsfxsize=8cm
\epsfbox{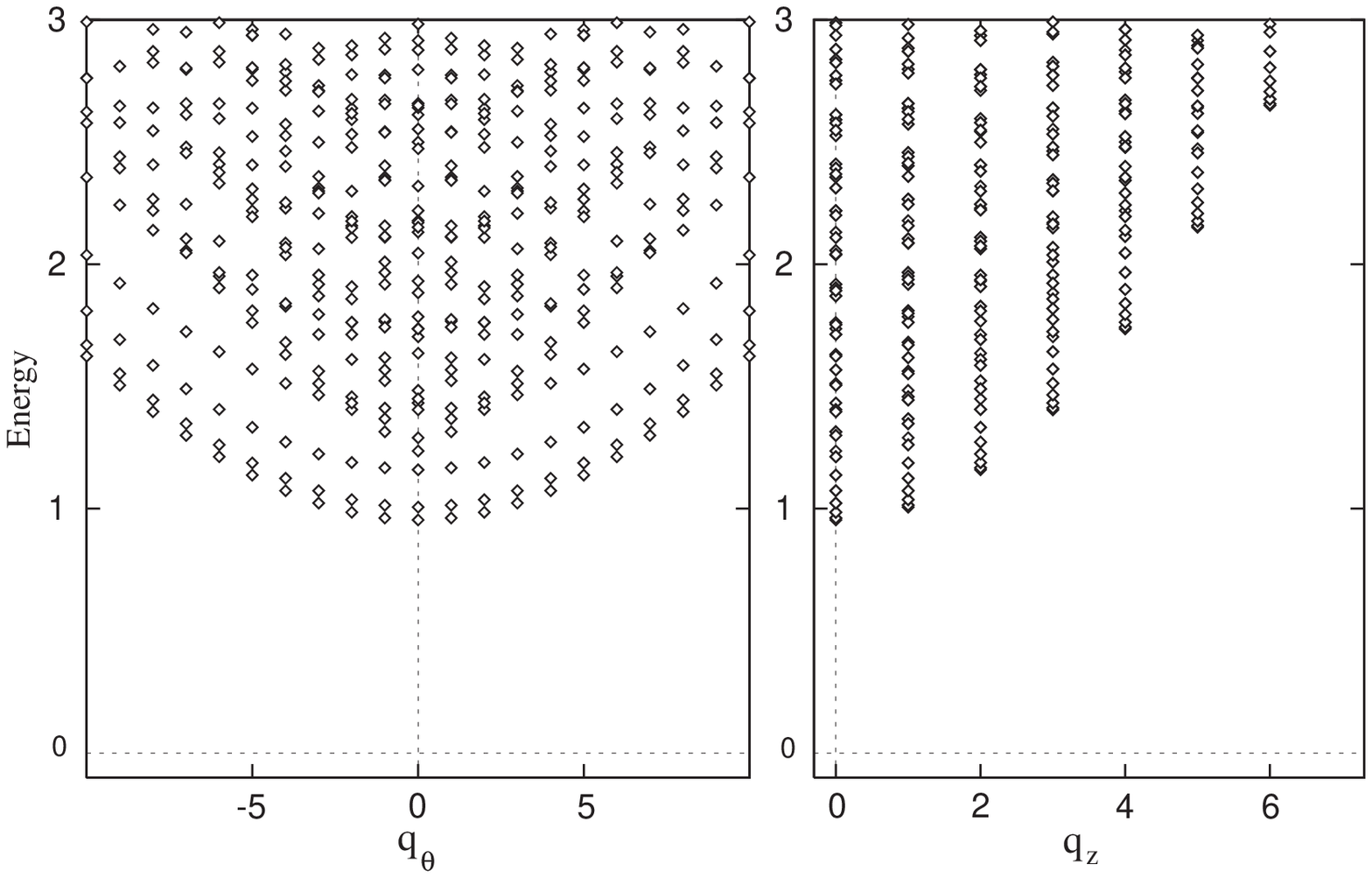}

\caption{
Excitation spectra as functions of $q_{\theta}$ (left hand side) and
$q_z$ (right hand side) in BA for the small wave numbers (a) and the overall
wave numbers at $q_{\theta}=0$ (b). The analytical expression eq.\ (35) of the
Bogoliubov spectrum in the long-wave length limit (straight line) is drawn in
the right hand figure of 4(a), and eq.\ (36) in the short wave length limit
(lower edge parabola) drawn in (b), showing  that they agree in those limits. 
(c) Excitation spectra as functions of $q_{\theta}$ (left hand side) and
$q_z$ (right hand side) in HFB. Note that it has a gap whose magnitude is
about 0.9.
} 
\end{figure}
%-------------------------------------------

In Fig.\ 5 we display the local density of states:
\begin{equation}
  N(E, r)=\sum_{\bf q} \{
    |u_{\bf q}(r)|^{2} +|v_{\bf q}(r)|^{2}
  \}\delta(E_{\bf q}-E)
\end{equation}
a combination of the quasi-particle eigenfunctions
with the lowest energies in BA (the results for the other approximations
are quite similar).
It is seen from this that although the weight 
spreads out to the entire regions, the major one 
concentrates near the wall boundary.

%-------------------------------------------
\begin{figure}
\epsfxsize=8cm
\epsfbox{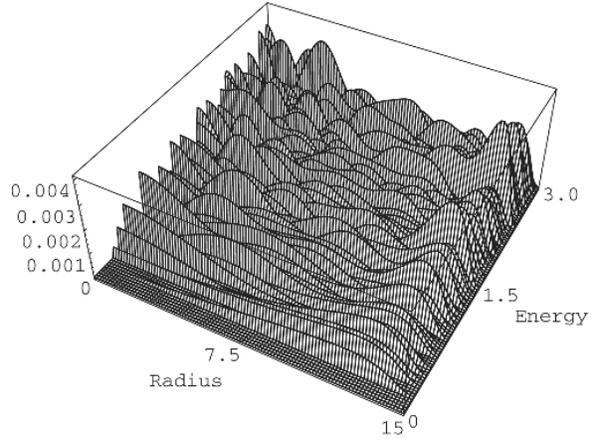}

\caption{
Local density of states $N(E,r)$ in the lower energy region for BA.
It is noted that the density of states accumulated at the center $r=0$ and
the wall $r=15$.
}
\end{figure}
%-------------------------------------------

\section{Rotating vortex case}
Having established the reliability of our 
numerical method,
we now discuss the results of the isolated singly quantized 
vortex case whose quantization unit is ${h\over m}$.
It is expected that a rotating BEC system whose frequency exceeds a certain
critical value
$\omega_c \sim {\hbar\over mR^3}\ln{R\over \xi}$ estimated as
$\sim 50$rad/s 
for $^{87}$Rb~\cite{fetterlt} sustains a quantized vortex line threading along
the cylindrically symmetric $z$-axis.

It quickly becomes clear after several numerical trials that
the BA and PA do not fulfill a fundamental requirement, that is,
the quasi-particle eigenvalues $E({\bf q})$ in eqs.\ (13) and (14) or 
eq.\ (27)
for BA ($\rho=0$ and $\Delta=0$) and also for PA ($\Delta=0$) must be positive
because the condensate situates at zero energy. 
The negative
% (or complex)
eigenvalue means 
an instability of the vortex state. Thus these approximations cannot be
a consistent theory for describing the vortex state.
As seen from Table 1 where the quasi-particle eigenvalues $E({\bf q})$ from
the lowest energies are listed for BA and HFB, 
the lowest eigenvalue with $q_{\theta}=-1$,
$q_z=0$ and $q_r=1$ in BA is negative.
We have checked the negative eigenvalue in BA by changing several conditions:
The system size ($R$, $L$), the interaction strength, and the Hamiltonian
matrix size. The negative eigenvalue belonging to the lowest energy in BA always
exists and is not an artifact of our numerical computations.
As for PA, the lowest eigenvalue with the same quantum number mentioned above
becomes also a negative number, which always appears
at every steps of the iteration processes for self-consistency.
We never complete a self-consistency and
cannot obtain a self-consistent solution.
We conclude that PA cannot sustain a stable vortex solution.
Thus, we are left with the non self-consistent GP and the full
self-consistent HFB which  are discussed in full details in the following.

%-------------------------------------------
\begin{figure}
\epsfxsize=8cm
\epsfbox{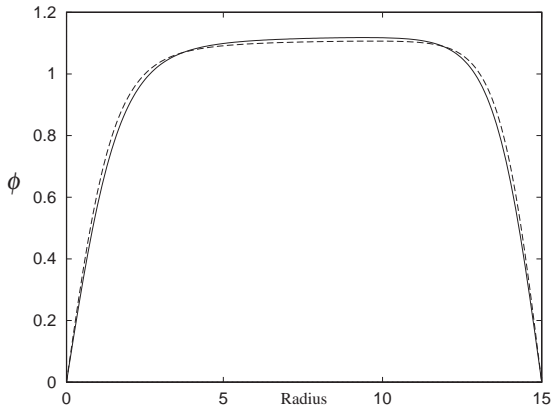}
\caption{
Spatial variation of the condensate $\phi(r)$ as a function of $r$ for
HFB (bold line) and GP (dashed line).
It is seen that at the vortex core $r=0$ it vanishes linearly, recovering
its bulk value within the characteristic length $\xi_a$.
}
\end{figure}
%-------------------------------------------

The spatial variations of the condensate $\phi(r)$ 
in HFB and GP are shown in Fig.\ 6. It is seen from this that 
both are almost identical, but the condensate is
pushed out in HFB by the presence of the non-condensate,
resulting in a slightly larger core radius. This could
be further amplified when the interaction $g$ becomes stronger
or the atomic density $n_a$ becomes high
since the non-condensate fraction increases.
It is also noticed that understandably $\phi(r)$ is almost 
symmetric at the middle $r=7.5$ because the characteristic length scale is
$\xi_a$ in this system, which governs the spatial variation at the core and wall.
At the vortex core $\phi(0)=0$ and linearly rises to recover its bulk
 value shown in Fig.\ 1. It will be interesting to
check if there is a similar Kramer-Pesch effect~\cite{kp} seen in
superconductors where the core radius shrinks as temperature decreases.
If indeed exists, the core radius increases further as temperature rises.
This temperature effect belongs to a future problem.
This should be checked experimentally in the present 
BEC systems because the expected core radius is far larger than that
in superfluid $^4$He.

%-------------------------------------------
\begin{figure}
\epsfxsize=8cm
\epsfbox{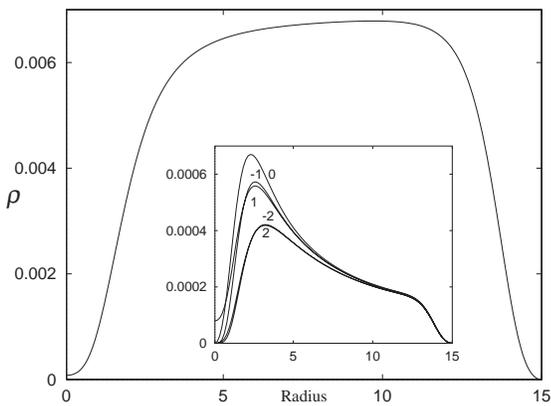}
\caption{
Spatial variation of the non-condensate $\rho(r)$
as a function of $r$ for HFB.
It is seen that $\rho(0)$ is non-vanishing,
coming from non-vanishing contribution with
$q_{\theta}=0$. The inset shows the contributions from
various components characterized by the quantum numbers:
$q_{\theta}=0$, $\pm1$, and $\pm2$. Note the different scales
of the vertical axes.
}
\end{figure}
%-------------------------------------------

The non-condensate $\rho(r)$ is displayed in Fig.\ 7.
According to Fetter~\cite{fetter},
a universal relation $\rho(0)/\rho(\infty)\sim$1.4 at $T=0$ is derived for BA,
independent of the 
interaction strength $g$, that is, the amplitude of $\rho(0)$ at
the core must exceed that in the outside region.
This prediction is not supported by the present calculation in HFB.
On the contrary, our result does show a suppression of $\rho(r)$ around the
core region whose characteristic length $\sim \xi_a$. We do not consider the
origin of this discrepancy further because BA is not a stable theory for
describing the vortex as mentioned before.
In the HFB result, which is a stable solution, $\rho(r)$  recovers the
bulk value$\sim$0.006 (see Fig.\ 2) far from the core where $\rho(0)$ reduces
to almost zero.
The characteristic recovery length is evidently longer than that in the
condensate as seen from Fig.\ 6.
This is partly because the behavior in $\rho(r)$ near the core is
quadratic in $r$ while that in $\phi(r)$ is linear.
The main contribution to the non-vanishing $\rho(0)$   
comes from the component with $q_{\theta}=1$
as seen from the inset of Fig.\ 7 where other dominant components
near the vortex core are also depicted.
This also explains the quadratic behavior in $r$ at the core mathematically.

%-------------------------------------------
\begin{figure}
\epsfxsize=8cm
\epsfbox{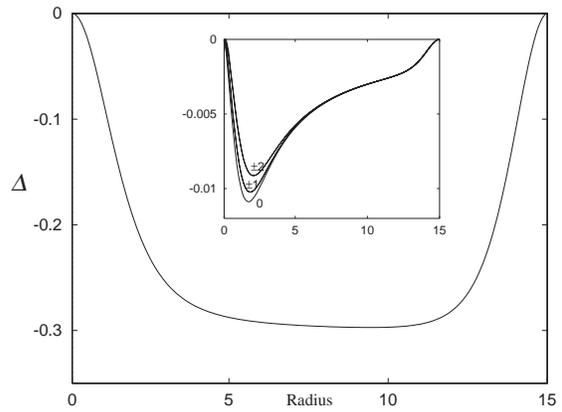}
\caption{
Spatial variation of the amomalous correlation $\Delta(r)$
as a function of $r$ for HFB.
The inset shows the contributions from various components
characterized by the 
quantum numbers:
$q_{\theta}=0$, $\pm1$, and $\pm2$. Note the different scales
of the vertial axes.
}
\end{figure}
%-------------------------------------------

The anomalous correlation $\Delta(r)$ is  shown in Fig.\ 8 for HFB,
which is no prior prediction and evaluated here for the first time.
The sign is negative same as in the non-vortex case of previous section and
$\Delta(r)$ vanishes quadratically 
at the core and also at the wall,
recovering its value ($\sim-0.3$) in the bulk (see Fig.\ 3 for comparison).
As shown as the inset where 
the contributions with the smaller $q_{\theta}$,
even near the vortex core there are no distinctive and/or dominant
contributions for $\Delta(r)$.
As mentioned in \S 3 this quantity strongly depends on the energy cutoff chosen.
If the cutoff energy increases, $|\Delta(r)|$ grows.
This feature is absent in the other quantities discussed here.

%-------------------------------------------
\begin{figure}
\epsfxsize=8cm
\epsfbox{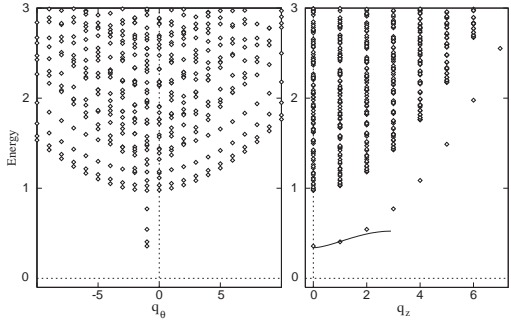}
\caption{
Excitation spectra as functions of $q_{\theta}$ (left hand side) and
$q_z$ (right hand side) in HFB.  
The distinctive modes at 
$q_{\theta}=-1$ corresponds to the Kelvin wave
whose dispersion relation eq.\ (38)
obtained by Pitaevskii~\cite{pitaevskii} in the long wave length is drawn in
the right hand side by adjusting the vertical axis for comparison.
The bulk of the continuum excitation spectra coincides with those in 
Fig.\ 4(c).
}
\end{figure}
%-------------------------------------------

In Fig.\ 9 we exhibit the excitation spectra as functions of $q_{\theta}$ and
$q_z$ where the distinctive excitations 
at $q_{\theta}=-1$ are seen, which are isolated from the rest of the continuum
seen before (Fig.\ 4(c)). This particular isolated excitation is known as the
Kelvin wave which is present in 
a vortex of classical liquid and corresponds to a helical mode of the
vortex line~\cite{donnelly}.
According to Pitaevkii~\cite{pitaevskii} who found it in BA,
at the long wave length limit the 
dispersion relation is given by 
\begin{eqnarray}
E({q_{\theta}=-1}, k_z)
={\hbar^2k^2_z\over 2m}\ln{1\over k_z\xi_a}\ \ \ (k_z\xi_a\ll 1).
\end{eqnarray}
Since this expression is valid only for an infinite system,
it is hard to judge whether or not our numerical result 
for a finite system agrees with this. 
Apparently, while the eq.\ (38) is gapless, the present 
result has a gap. Apart from the gap, our result for the dispersion relation 
does not contradict
this behavior (see the line of eq.\ (38) drawn Fig.9).
The bulk of the gapful continuum in Fig.\ 9 is
just the same as in Fig.\ 4(c) as expected.

The local density of states given by eq.\ (37) 
with the lower energy side are shown in Fig.\ 10 where the states
distinctively localized near the core correspond
to the angular momentum $q_{\theta}=-1$.
The detailed analyses of eqs.\ (13) and (14)
for $q_{\theta}=-1$ and $q_{\theta}=1$ in BA are performed by
Pitaevskii~\cite{pitaevskii} and Iordanskii~\cite{iordanskii}
respectively: $u_{q_{\theta},q_z}(r)\sim r^{|q_{\theta}+1|}(1+O(r^2))$ 
and $v_{q_{\theta},q_z}(r)\sim r^{|q_{\theta}-1|}(1+O(r^2))$
which are easily derived by analyzing eqs.\ (13) and (14) for BA.
Note that only $u_{q_{\theta=-1},q_z}(0)$ and $v_{q_{\theta=1},q_z}(0)$
are non-vanishing at the core.
These properties are also true for the full self-consistent HFB.
The local density of states in  superconductors is directly observed
by scanning tunneling microscope~\cite{hess} and analyzed theoretically
within the similar theoretical framework~\cite{gygi} quite successfully.
Since as mentioned before, the core radius in BEC systems 
is relatively large, there is good chance to directly observe
the local excitation spectrum possibly by an optical method
once the vortex can be created.

%-------------------------------------------
\begin{figure}
\epsfxsize=8cm
\epsfbox{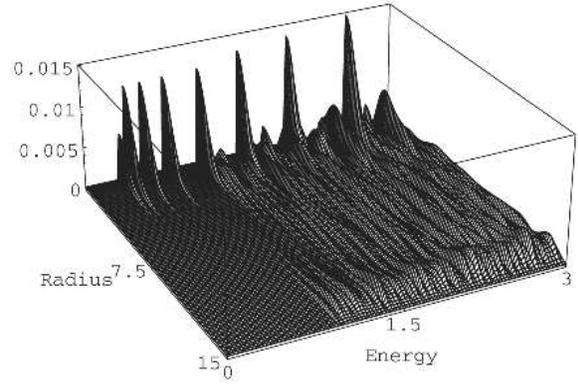}
\caption{
Local density of states $N(E_q,r)$ in the lower energy region for HFB.
It is noted that the density of states accumulates at the center $r=0$ and
the wall $r=15$.
The distinctive peak structures come from those with 
the quantum number $q_{\theta}=-1$, indicating that these modes
localize at the vortex core.
}
\end{figure}
%-------------------------------------------

%-------------------------------------------
\begin{figure}
\epsfxsize=8cm
\epsfbox{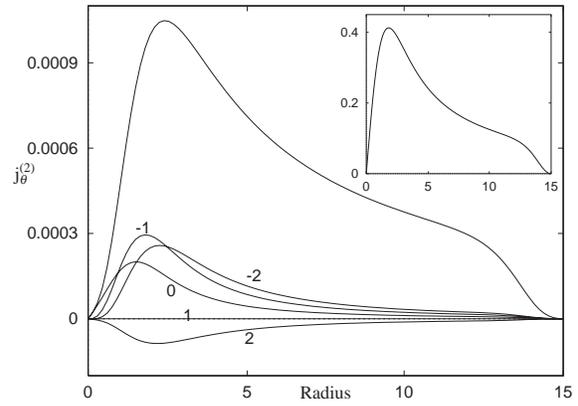}
\caption{
Current distribution $j^{(2)}_{\theta}(r)$ which comes from
the non-condensate where the contributions from the components with
$q_{\theta}=-0$,$\pm1$, and $\pm2$ are shown.
The inset shows the total current $j_{\theta}(r)$.
Note that the dominant contribution comes from the condensate.
The current is normalized by ${\hbar\over m}\cdot{n_a\over \xi_a}$.
}
\end{figure}
%-------------------------------------------

Finally the circulating current density $j_{\theta}(r)$ for
the $\theta$-component which is expressed as 
\begin{eqnarray}
  j_{\theta}(r)&=&j^{(1)}_{\theta}(r)+j^{(2)}_{\theta}(r)
\\
  j^{(1)}_{\theta}(r)&=&{\hbar\over m}{n_a\over  \xi_a}\cdot{\phi^2(r)\over r}
\\
  j^{(2)}_{\theta}(r)&=&
  -{1\over 2{\rm i}r}\cdot{\hbar\over m}{n_a\over \xi_a} \nonumber
\\&&
  \sum_{\bf q}
  \left\{
    v^{\ast}_{\bf q}(r){\partial v_{\bf q}(r)\over \partial\theta}
    -v_{\bf q}(r){\partial v^{\ast}_{\bf q}(r)\over \partial\theta}
  \right\}
\end{eqnarray}
where the total current consists of the condensate component
$j^{(1)}_{\theta}(r)$ and the non-condensate $j^{(2)}_{\theta}(r)$.
The non-condensate contribution $j^{(2)}_{\theta}(r)$
normalized by ${\hbar\over m}\cdot{n_a\over  \xi_a}$
is depicted for HFB in Fig.\ 11 where each with $q_{\theta}=0,\pm1,\pm2$ is 
shown separately,
and those with smaller $|q_{\theta}|$ dominate near the core. 
As is seen from Fig.\ 11 $j^{(2)}_{\theta}(r)$ in the immediate 
vicinity of the core is governed by the $q_{\theta}=0$
component. The negative (positive) $q_{\theta}$'s give rise
to the positive (negative) contribution  to $j_{\theta}(r)$.
The inset shows the total current density $j_{\theta}(r)$ where 
$j_{\theta}(r)\sim r$ for small $r$ and 
$j_{\theta}(r)\sim {1\over r}$ for larger $r$.
The relative weights of $j^{(1)}_{\theta}(r)$ and $j^{(2)}_{\theta}(r)$ depends
on the interaction strength and the average density $n_a$. 

\section{Conclusion and Discussions}

We have investigated various approximate mean-field type
theories (Gross-Pitaevskii, Bogoliubov theory, Popov theory and
Hartree-Fock Bogoliubov theory) within the framework of the Bogoliubov
approximation for a dilute Bose gas, on which renewed
interest  is focused recently by the discovery of Bose-Einstein 
condensation in alkali atom gases.
The above four type theories are numerically solved for parameters appropriate
to ongoing experiments on $^{23}$Na and $^{87}$Rb atoms and analyzed
on an equal footing for the first time. We extract several properties of
experimental
interest in BEC systems, namely, the spatial structures of the condensate,
non-condensate and anomalous correlation both in stationary non-uniform case
and the vortex case under rotation.
A numerical procedure for solving these mean-field equations are
presented and critically assessed for future use.

In the case of the stationary non-uniform Bose gas
confined in a cylindrically symmetric vessel 
the above four theories yield almost identical results
for the spatial profile of the condensate.
The non self-consistent BA and self-consistent PA and HFB
give  similar profiles for the non-condensate,
but in the last the magnitude are halved.

In the vortex case these mean-field theories are 
numerically examined. It is found that GP and PA do not fulfill the
fundamental requirement, showing an instability
of the theories, and thus are inadequate for describing a vortex.
The full self-consistent solution for HFB is obtained and analyzed in detail.
The spatial structures of the vortex core
for Bose  systems;
the condensate, non-condensate and anomalous correlation
are explicitly derived for the first time.
Some characteristics of the local density of states
and circulating current are pointed out in the hope to be observed
in BEC systems  in alkali atom gases.

%%%%%%%%%%%%%%%%%%%%%%%%%%%%%%%%%%%%%%%%%%%%%%%%%%%%%%%%%%%%%%%%%%%%%%
\section*{Acknowledgments}
The authors thank M. Ichioka, N. Enomoto, and N. Hayashi for useful discussions.

%----------------------------------------

\begin{table}
\begin{tabular}{@{\hspace{\tabcolsep}\extracolsep{\fill}}cccccccc}
\hline
   $q_\theta$ & $q_z$ & $q_r$ & Bogoliubov & $q_\theta$ & $q_z$ & $q_r$ & HFB \\
\hline
   -1  &  0  &  1  & -0.01034608886000 & -1  &  0  &  1  &  0.35866592764172 \\
    0  &  0  &  1  &  0.00222477348689 & -1  &  1  &  1  &  0.40521222215704 \\
   -1  &  1  &  1  &  0.08359597874556 & -1  &  2  &  1  &  0.54360716036381 \\
   -1  &  0  &  2  &  0.20124587759088 & -1  &  3  &  1  &  0.77139318149905 \\
    1  &  0  &  1  &  0.20342514914009 & -1  &  0  &  2  &  0.97584202447130 \\
   -1  &  2  &  1  &  0.26341740112590 &  0  &  0  &  1  &  0.98055487186421 \\
   -2  &  0  &  1  &  0.28575943668855 & -2  &  0  &  1  &  0.98631908786634 \\
    0  &  1  &  1  &  0.31282158360176 &  1  &  0  &  1  &  1.00123548991620 \\
    2  &  0  &  1  &  0.35670360464514 & -3  &  0  &  1  &  1.01170602030422 \\
   -1  &  1  &  2  &  0.35731592142005 & -1  &  1  &  2  &  1.02820298664511 \\
    1  &  1  &  1  &  0.38616772030053 &  0  &  1  &  1  &  1.03306239243685 \\
   -3  &  0  &  1  &  0.39673767766612 &  2  &  0  &  1  &  1.03656925663829 \\
\hline
\end{tabular}
\caption{
The lowest energies for the Bogoliubov approximation
and the Hartree-Fock Bogoliubov theory (HFB).
}
\end{table}

\end{document}